Global poverty · a statistical analysis

# Global Poverty Beyond the Official Line

## *A Bounded Estimate of Material Insufficiency*

An integrative synthesis of monetary, multidimensional, and relative deprivation measures

How many people are poor? It depends on the line drawn.

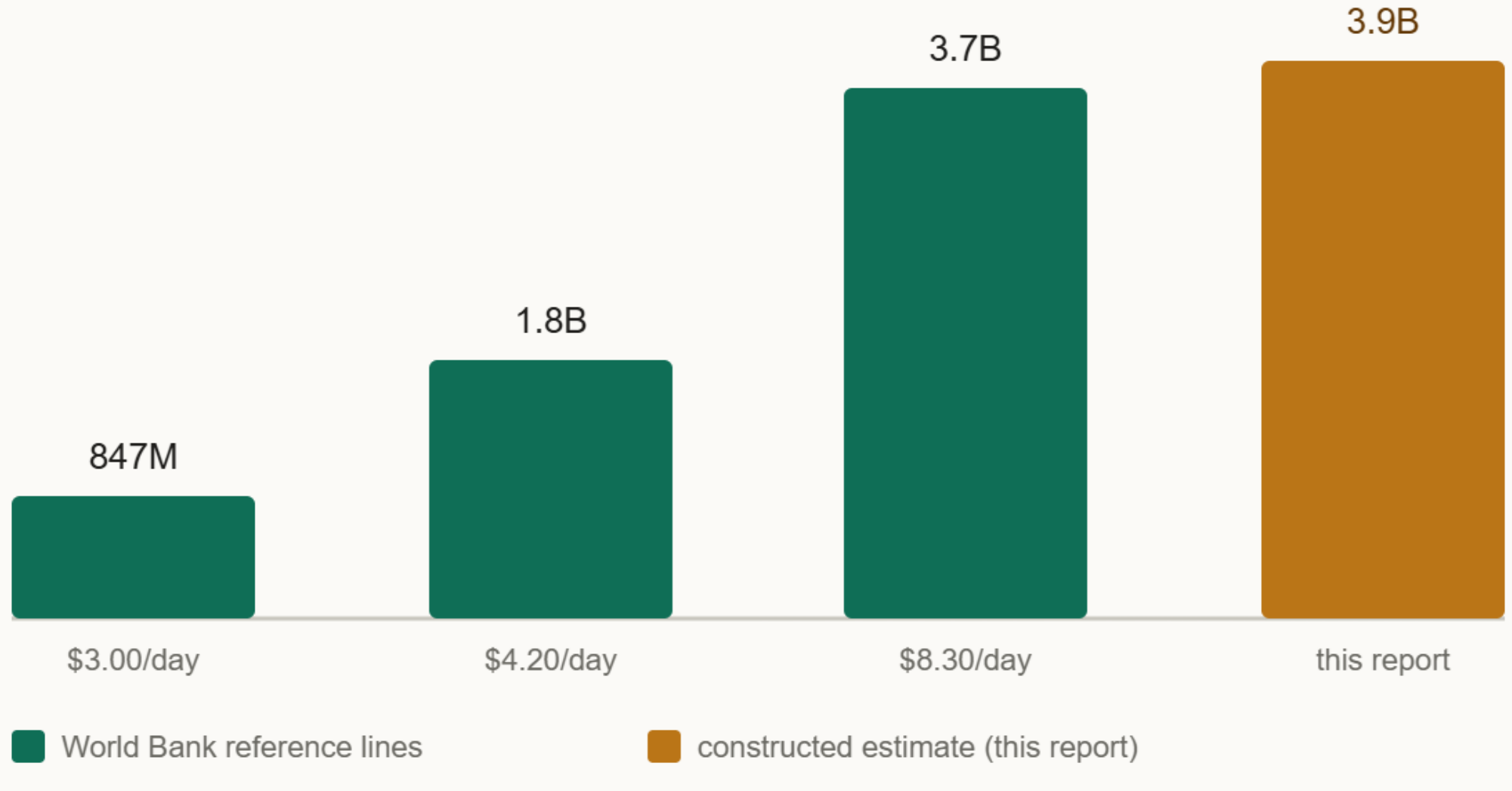

Giancarlo Crocetti
Assistant Professor & Director, AI Institute
crocettg@stjohns.edu
Editorial Team · Institute for Machine Learning, Danbury, CT

## Recommended citation

Crocetti, G., & Editorial Team (2026). Global Poverty Beyond the Official Line: A Bounded Estimate of Material Insufficiency. Institute for Machine Learning, in collaboration with the AI Institute, St. John's University.

## Copyright and license

## Contact

Giancarlo Crocetti, corresponding author
crocettg@stjohns.edu
AI Institute, St. John's University, Queens, NY 11439
Institute for Machine Learning, Danbury, CT 06811

## Disclaimer

The findings, interpretations, and conclusions in this report are those of the authors and do not necessarily represent the views of the Institute for Machine Learning, St. John's University, or the organizations whose data are cited. The central estimate of “material insufficiency” is an analytical construct, stated with its assumptions in Section 13, and is not an official poverty count.

## Data and sources

All figures derive from publicly available institutional sources — the World Bank, UNDP/OPHI, FAO, ILO, OECD, Eurostat, and national statistical agencies — listed in full in the report's References. Data are current through March 2026.

# Contents

## List of Figures

## List of Tables

| Abbreviation | Full form |
|---|---|
| CFPB | Consumer Financial Protection Bureau (United States) |
| CPI | Consumer Price Index |
| EU | European Union |
| Eurostat | Statistical office of the European Union |
| FAO | Food and Agriculture Organization of the United Nations |
| FRED | Federal Reserve Economic Data (Federal Reserve Bank of St. Louis) |
| FY | Fiscal year |
| GDP | Gross domestic product |
| H1 | First half of the (fiscal) year |
| ILO | International Labour Organization |
| IPL | International Poverty Line |
| ISTAT | Istituto Nazionale di Statistica (Italy's national statistics institute) |
| LAC | Latin America and the Caribbean |
| LMIC | Lower-middle-income country |
| MENA | Middle East and North Africa |
| MHLW | Ministry of Health, Labour and Welfare (Japan) |
| MPI | Multidimensional Poverty Index (Global MPI) |
| NITI Aayog | National Institution for Transforming India |
| OECD | Organisation for Economic Co-operation and Development |
| OPHI | Oxford Poverty and Human Development Initiative |
| OPM | Official Poverty Measure (United States) |
| PIP | Poverty and Inequality Platform (World Bank) |
| PPP | Purchasing power parity |
| SDG | Sustainable Development Goal |
| SOFI | The State of Food Security and Nutrition in the World (FAO) |
| SPM | Supplemental Poverty Measure (United States) |
| UN | United Nations |
| UNDP | United Nations Development Programme |
| USDA | United States Department of Agriculture |
| WDI | World Development Indicators (World Bank) |
| WHO | World Health Organization |

## Acknowledgments and Disclosures

**Acknowledgments.** This report rests on the public data infrastructure maintained by the World Bank, UNDP and OPHI, the FAO, the ILO, the OECD, Eurostat, and the national statistical agencies whose household surveys underpin the figures presented here; their commitment to open data made this synthesis possible.

**Funding.** This research received no external funding. It was undertaken by the Institute for Machine Learning as independent research.

**Competing interests.** The authors declare no competing financial or personal interests.

**Editorial independence.** The Institute for Machine Learning is both a contributing author and the publisher of this report. Its analysis draws solely on publicly available data, and the findings, interpretations, and recommendations are the independent conclusions of the authors and are not subject to external or institutional approval of the content.

**Use of generative AI.** In keeping with the shared commitment stated at the front of this report, its preparation deliberately used generative artificial intelligence as both a working tool and a demonstration of how such technology can serve the public interest. AI assistance contributed to drafting, structuring, and synthesizing the cited sources, data visualization, and editing under the direction of the authors. The authors reviewed and edited all content, verified the figures and citations against their primary sources, and are solely responsible for the report's analysis, accuracy, and conclusions.

# Global Poverty Beyond the Official Line: A Bounded Estimate of Material Insufficiency

*An integrative synthesis of monetary, multidimensional, and relative deprivation measures*

**Giancarlo Crocetti[1]* · Editorial Team[2]**

[1] Assistant Professor and Director, AI Institute, St. John's University, Queens, NY 11439
[2] Institute for Machine Learning, Danbury, CT 06811
* Corresponding author: crocettg@stjohns.edu

## Executive Summary

Numbers this large invite a defensive reflex: reach for the reassuring figure and move on. By the most widely cited measure — the World Bank's extreme-poverty line of $3.00 per day (2021 PPP) — approximately 847 million people, or 10.4% of the world's population, lived in poverty in 2024. That figure is accurate as measured. It is also, by design, a floor: a threshold built to mark bare survival in the poorest economies, not to describe what it takes for a family anywhere to live with basic security.

The central argument of this report is that the reassurance offered by that single line is largely an artifact of how we chose to measure. Held to the $3.00 floor, roughly 847 million people are poor; held to the standards of their own societies, the number is several times larger, and even the most cautious, honest count exceeds a billion. Measured against income standards appropriate to each country's level of development, some 3.7 billion people fall below the World Bank's $8.30/day upper-middle-income reference line, roughly 1.1 billion are multidimensionally poor, and around 2.3 billion are food insecure. These are not competing errors but answers to different questions.

This report does not dispute the official data. It takes those figures as accurate and asks what they do and do not capture. It disaggregates by geography, age, and gender and supplements the monetary count with multidimensional measures of food security, working poverty, social protection, and relative poverty. It also documents a data-quality problem that runs in one direction: survey coverage is weaker in regions with the worst poverty, so official figures are far likelier to understate deprivation than to overstate it.

Where the report advances a figure of its own, it labels it as a construct rather than a finding. Under an explicit standard — the inability to meet basic needs as defined by one's own society — a bounded central estimate places approximately 3.9 billion people, close to half of humanity, in material insufficiency. This is not a World Bank poverty count and is not presented as one; it is one defensible way to answer a question the official line does not address, reported with its assumptions and a wide uncertainty band. The honest answer is a range — from a conservative floor above one billion to an upper bound approaching five — and the width of that range is itself the finding.

Behind the range are people the headline renders invisible: roughly half of those in extreme poverty are children, who did not choose their circumstances; hundreds of millions of adults who work full time and still cannot feed their families; and, in the wealthiest economies on earth, children in food-insecure homes and parents one medical bill from ruin. These households are not passive — they navigate deprivation with work, planning, and networks of mutual support — but effort cannot substitute for a wage floor or a safety net that is not there. The report asks the reader not to accept a single figure, but to see that the most familiar number answers a far narrower question than it appears to, and to hold the fuller picture, uncertainty and all.

*Data through March 2026. Scope and method: an integrative synthesis of secondary institutional data (World Bank, UNDP/OPHI, FAO, ILO, OECD, Eurostat, and national statistical agencies), not a systematic review or meta-analysis. Estimates are reported as ranges with graded certainty; principal limitations are cross-vintage PPP comparability, imputation error in conflict-affected states, and overlap between measures.*

## Key Messages

- **The most-cited figure answers a narrower question than it appears to.** Around 847 million people (10.4%) fall below the World Bank's extreme-poverty line—a bare-survival floor calibrated for the poorest economies. It is accurate, and it is not a measure of how many people are poor.
- **The count is standard-dependent, and the range is enormous.** The same data yield ~847M at $3.00/day, ~1.1B in multidimensional poverty, and ~3.7B below the $8.30/day line. No single number is meaningful without its threshold.
- **On a defensible standard, close to half of humanity lives in material insufficiency.** Measured against the norms of one's own society, a bounded central estimate is ~3.9 billion. This is an analytical construct, stated with its assumptions, not a World Bank poverty count.
- **Measurement is weakest where poverty is worst, so the error runs one way.** Fewer than half of countries have post-2020 survey data, and the highest-poverty regions have the thinnest coverage. Official figures are far likelier to undercount than to overcount.
- **Deprivation in rich countries is real and invisible to every international line.** Food-insecure children, working single parents below national poverty lines, and households in energy or medical-debt distress are measurable by those countries' own standards and absent from every global count.

*Table 1. Key figures at a glance*

| Metric | Figure | Definition/threshold | Type | Source |
|---|---|---|---|---|
| Extreme poverty | ~847M (10.4%) | $3.00/day, 2021 PPP | Official | World Bank PIP, Mar 2026 |
| Multidimensional poverty | 1.1B (13.5%) | MPI acute deprivation | Official | UNDP/OPHI, 2024 |
| Below $8.30/day | ~3.7B (45.5%) | Upper-middle-income reference line | Official (reference line) | World Bank, Fall 2025 |
| Food insecure | 2.3B (28.0%) | Moderate + severe | Official | FAO SOFI, 2025 |
| Without social protection | 3.8B (47.2%) | No benefit coverage | Official | ILO / UN SDG, 2023 |
| Material insufficiency (central estimate) | ~3.9B (~47%) | Basic needs vs. one's own society norms | Constructed (this report) | Author's estimate, §13 |

The final row is an analytical construct, not an official statistic; the distinction is developed in Section 13.

**Framing note.** This report takes official data as accurate and asks what those data do and do not capture. Where it advances an interpretation beyond the source figures, most importantly, the central estimate in Section 13, that interpretation is labeled as a construct rather than a finding.

## 1. Methods, Scope, and Certainty Framework

**Question.** How many people cannot reliably meet basic needs, and how sensitive is the answer to the measurement standard applied?

**Key terms.**

- *Extreme poverty* — living below $3.00/day (2021 PPP), the World Bank's floor calibrated to the poorest economies.
- *Multidimensional poverty* — deprived in at least a third of the MPI's weighted health, education, and living-standard indicators.
- *Relative poverty* — living below 50% of a country's median disposable income; a within-society standard used chiefly in high-income countries.
- *Near-poverty / vulnerable* — above the poverty line but within one shock of falling below it (for example, the $6.85–$14/day band).
- *Material insufficiency* — this report's construct: the inability to meet basic needs by the standards of one's own society.
- *PPP (purchasing power parity)* — a currency conversion that equalizes what money buys across countries; every line is anchored to a base year (here 2017 or 2021).

**Sources and selection.** Evidence is drawn from primary institutional releases and peer-reviewed literature. Selection was purposive rather than systematic: for each dimension, the most authoritative and most recent source offering global or comparable cross-national coverage was prioritized. No formal search protocol was used, which is a limitation relative to systematic review standards.

**Threshold and PPP conventions.** Unless stated otherwise, monetary figures use the 2021 PPP reference lines ($3.00 / $4.20 / $8.30 per day). Where only 2017 PPP figures ($2.15, $3.65) were available, this is flagged inline. Figures on different PPP bases are not directly comparable and are never combined.

**Treatment of overlapping measures.** The dimensions reported, monetary, multidimensional, food insecurity, working poverty, social protection, and relative poverty, overlap substantially and are not summed, with the sole exception of the explicitly labeled composite estimates in Section 13. In those estimates, populations are combined only where they are disjoint: specifically, residents of high-income countries who fall below a national relative line and above $8.30/day in PPP terms are added to the $8.30 count, because those two sets do not intersect.

**Certainty ratings.** Each headline figure is graded High, Medium, or Low against four criteria: source authority, data recency (post-2020 preferred), population coverage (share measured versus modeled), and methodological transparency. High denotes an authoritative, recent, high-coverage figure derived by a transparent method; Low denotes a figure that is extrapolated, dated, or based on low survey coverage.

**Handling of stale and modeled data.** Country figures resting on pre-2020 surveys are marked "stale"; modeled or imputed figures are identified as such. Because survey coverage is inversely correlated with deprivation, the expected direction of error is undercounting.

## 2. Data-Quality Constraints

An honest report discloses the quality of its evidence before presenting any statistic. Global poverty figures depend on household surveys, and for much of the world, particularly its poorest regions, those surveys are inadequate (World Bank PIP Coverage Blog, November 2024):

- Fewer than half of all countries have household survey data from 2020 or later.
- Sub-Saharan Africa, which holds 71% of all the extreme poor, had survey coverage of only 56.7% of its population as of September 2025. The remainder is modeled rather than measured.
- Pakistan's latest PIP survey dates from 2018, before a severe economic crisis. The national poverty rate reached 25.3% in FY2024, up seven percentage points from 2023, with roughly 13 million additional people falling into poverty (World Bank Pakistan projections; World Bank South Asia Blog, September 2025). At the lower-middle-income line, the rate reached 40.5% ($3.65/day, 2017 PPP — not directly comparable to the 2021 PPP lines used elsewhere). Headline consumer-price inflation ran at roughly 28.8% in the

first half of FY24 and averaged 29.2% across FY23 (World Bank Pakistan Development Update, April 2024; IMF Article IV, September 2024) — substantially higher than PIP-modeled figures would suggest.

- India had no nationally representative consumption survey between 2011 and 2022, an eleven-year gap during which the world's most populous country's poverty was modeled rather than observed.
- Eritrea's latest survey dates from 1993; Sudan's from 2009, well before the current civil war.

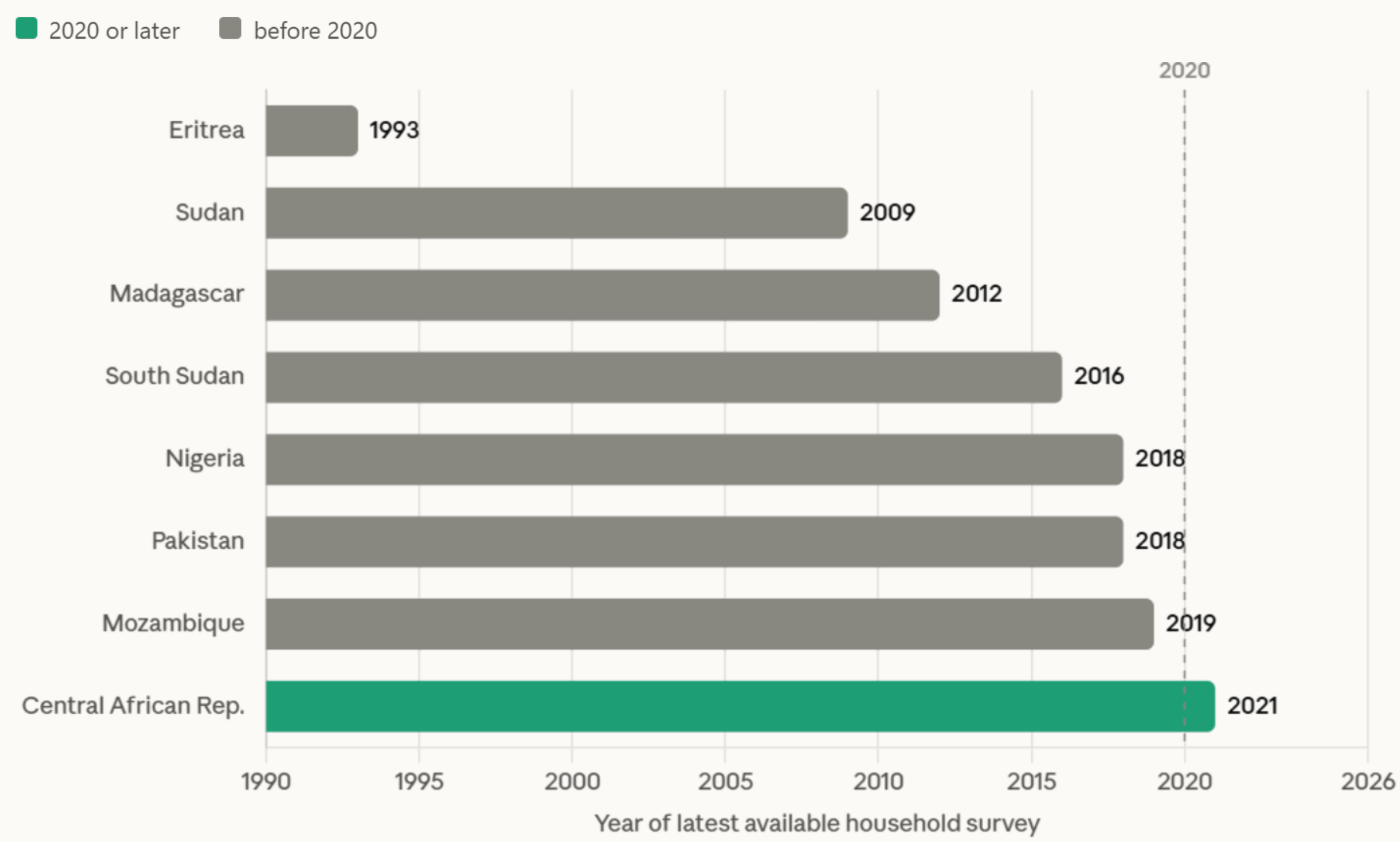

***Figure 1. Measurement is weakest where poverty is worst.***
*Sources: World Bank Poverty and Inequality Platform; World Bank PIP Coverage Blog (November 2024).*

The technique used to fill these gaps — survey-to-survey imputation — can produce seriously distorted results when underlying conditions have changed (World Bank / Universidad de Los Andes, August 2025). In conflict-affected countries and those in economic crisis, imputed estimates may understate poverty by a large, unknowable margin. The practical implication is that the countries most in need of accurate data are the least likely to have it, and that the true global count is likely higher, not lower, than official figures indicate.

## 3. Why a Single Line Is Insufficient

A poverty line answers a narrow question — is a person's income above a specified value — and cannot answer the one that matters to a family: can we live with stability and basic security? The International Poverty Line was revised in June 2025 from $2.15/day (2017 PPP) to $3.00/day (2021 PPP), reflecting updated cost-of-living data from more than 160 countries (World Bank PIP, June 2025). The revision improved accuracy, but $3.00/day is roughly $1,095 per year, a sum that does not reliably cover rent, food, healthcare, and clothing in virtually any urban area.

The World Bank maintains three reference lines, each calibrated to a different income context:

***Table 2. World Bank reference poverty lines and the population below each, 2024.***

| Reference line | 2021 PPP | Applicable to | People below (2024) |
|---|---|---|---|
| Extreme poverty | $3.00/day | Low-income countries | ~847M (10.4%) |
| Lower-middle threshold | $4.20/day | Lower-middle-income countries | ~1.8B (approximate; no verified global aggregate) |
| Upper-middle threshold | $8.30/day | Upper-middle-income countries | ~3.7B (45.5%) |

Two methodological points govern how these figures are used in this report. First, the $8.30/day figure is the reference threshold for upper-middle-income countries — Brazil, China, South Africa, Mexico, Thailand, and comparable economies — not a universal global poverty line. The World Bank does not itself characterize all 3.7 billion people below $8.30/day as "poor" in a single unified sense; it reports the global count below each reference line to enable cross-country comparison.

The defensible statement is therefore that 3.7 billion people live below the income standard most appropriate for measuring poverty in upper-middle-income contexts. For low-income countries, the $3.00/day line is the more appropriate reference; for high-income countries, even $8.30/day understates deprivation. Section 13 makes explicit how this report uses the $8.30 line as an analytical anchor rather than treating it as a finding.

Second, the $4.20/day figure is labeled approximate because no verified global aggregate for this threshold has been published in consolidated form under the revised 2021 PPP lines. India alone accounts for roughly 341 million people at this line; the ~1.8 billion estimate is reasonable but should be read as an approximation.

## 4. Extreme Poverty: The Official Count

The September 2025 update of the Poverty and Inequality Platform placed the 2024 global extreme-poverty count at 839 million (World Bank, September 2025), an increase of roughly 22 million over the prior vintage, driven primarily by new survey data from Nigeria confirming that extreme poverty in Western and Central Africa is worse than previously estimated. The March 2026 update revised the global share to 10.4%, placing the absolute number at approximately 847 million (World Bank PIP, March 2026). The 2025 nowcast — a model-based projection rather than a survey-based count, and subject to revision in later vintages — is approximately 808 million (10.1%).

The extremely poor are concentrated and young. More than three-quarters live in Sub-Saharan Africa or in fragile and conflict-affected states (UN SDG Report, 2025). Some 451 million live in fragile or conflict-affected situations, where poverty is five times more prevalent than elsewhere (World Bank, Fall 2025). About 412 million are children aged 17 or younger — just under half of all people in extreme poverty, despite children being roughly 30% of the world's population. More than 400 million women face a 27–42% higher poverty risk than men, depending on the region.

SDG Target 1.1 — eliminating extreme poverty by 2030 — is highly unlikely on current trajectories. Projections show 8.9% of the world will still be in extreme poverty in 2030, requiring a roughly threefold acceleration in the rate of reduction; only one country in five is on track to halve its national poverty rate by then (UN SDG Report, 2025).

## 5. The Geography of Poverty

**Sub-Saharan Africa.** The region accounts for 71% of the world's extremely poor, yet holds only 18% of the world's population (World Bank, Fall 2025). Its extreme-poverty rate is 46.0% — nearly one in two people. Country-level extremes illustrate both the severity and the data problem:

***Table 3. The highest national poverty rates in Sub-Saharan Africa, by survey year***.

| Country | National poverty rate | Data year |
|---|---|---|
| South Sudan | 82.3% | 2016 (stale) |
| Madagascar | 75.2% | 2012 (stale) |
| Central African Republic | 68.8% | 2021 |
| Mozambique | 65.0% | 2019 |
| Nigeria | 56.2% | 2018 |

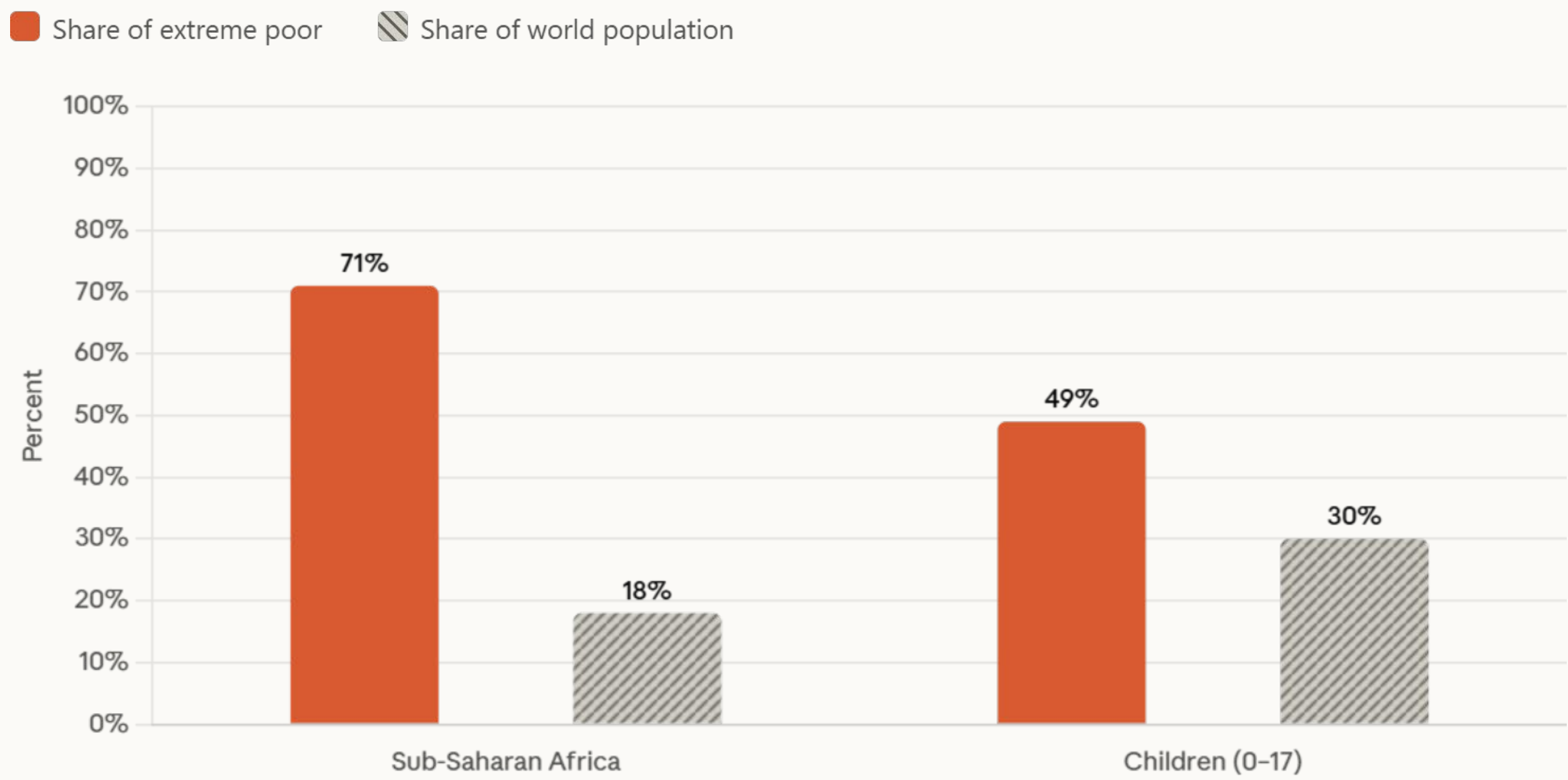

***Figure 2. Poverty is disproportionately concentrated by region and age.***
*Source: World Bank Fall 2025 Poverty and Inequality Update; UN SDG Report (2025).*

The "data year" column is not incidental. For South Sudan and Madagascar, the figures rest on surveys more than a decade old; given intervening conflict and climate disruption, current rates may differ substantially. Some 312 million children in poverty in the region represent 75% of the global total of children in extreme poverty, despite the region holding only about 23% of the world's children — an intergenerational crisis in which children born poor face a near-90% probability of spending their entire childhoods in poverty.

**South Asia and the India measurement debate.** South Asia has made real progress; child poverty in the region has more than halved between 2014 and 2024. India illustrates how profoundly the choice of measurement affects the count. Following its 2022–23 Household Consumption Expenditure Survey — the first nationally representative survey in over a decade — the World Bank revised India's extreme-poverty rate at $3.00/day (2021 PPP) to 5.3%, about 75 million people. Credible Indian scholars have questioned this figure: Rangarajan and Dev (2024) estimate 10.8% on the Tendulkar methodology; Himanshu, Lanjouw, and Schirmer (2024) a 9.9–12.2% range; Sethu, Surya, and Ruthu (2024) 26.4% on the Rangarajan Committee methodology. The analytically appropriate line for India, a lower-middle-income country, is the $4.20/day threshold, which yields about 24% of the population, or roughly 341 million people. India's own Multidimensional Poverty Index (NITI Aayog, 2023) placed non-monetary poverty at 11.28% in 2022–23, down from 29.17% in 2013–14.

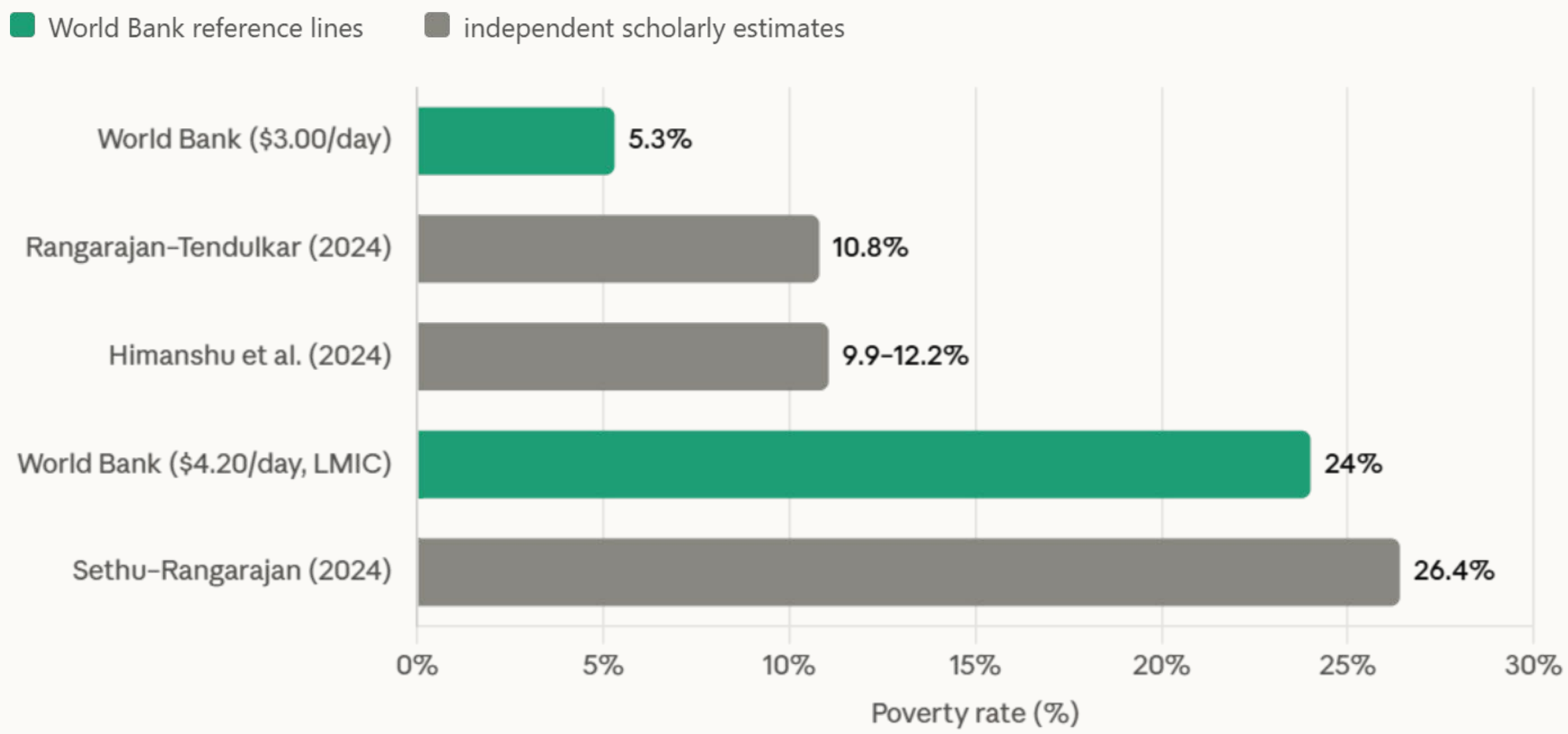

***Figure 3. One country, many answers: India's poverty rate by method.***
*Sources: World Bank PIP; Rangarajan and Dev (2024); Himanshu, Lanjouw, and Schirmer (2024); Sethu, Surya, and Ruthu (2024).*

The lesson is consequential: the choice of threshold in India alone swings the global count by hundreds of millions. At $3.00/day, India contributes 75 million; at the LMIC-appropriate $4.20/day, 341 million.

**The hidden poor in middle-income countries.** About 64.5% of the 1.1 billion multidimensionally poor — roughly 740 million people — live in middle-income countries, not the poorest nations (UNDP MPI, 2024). Large, poor populations in Nigeria, India, Pakistan, Indonesia, Egypt, and Brazil are frequently invisible in global debates because their national income levels exclude them from extreme-poverty headlines.

## 6. Multidimensional Poverty

The Global Multidimensional Poverty Index (UNDP/OPHI) measures poverty across three dimensions — health, education, and living standards — and ten weighted indicators; a person is multidimensionally poor if deprived in at least one-third of the weighted indicators. By this measure, 1.1 billion people across 112 countries live in acute multidimensional poverty, about a third more than the extreme monetary count (UNDP MPI, 2024).

***Table 4. Dimensions and indicators of the Global Multidimensional Poverty Index.***

| Dimension | Indicators |
|---|---|
| Health | Nutrition; child mortality |
| Education | Years of schooling; school attendance |
| Living standards | Cooking fuel; sanitation; drinking water; electricity; housing; assets |

More than half of the 1.1 billion — 584 million — are children. A child in multidimensional poverty is simultaneously exposed to malnutrition, school absence, indoor air pollution from solid-fuel cooking, and inadequate sanitation. These deprivations compound one another biologically and educationally in ways single-indicator measures do not capture. Monetary and multidimensional poverty overlap only partially: millions living above $3/day remain deprived across multiple non-monetary dimensions.

**Intra-household deprivation.** All monetary statistics measure deprivation at the household level, assuming equal resource-sharing. Research by World Bank economists (De Vreyer and Lambert) finds that roughly one in ten households classified as non-poor contains a poor member — typically a woman or child receiving less than the household average. All global counts are therefore structural undercounts of individual deprivation.

## 7. Near-Poverty and Deprivation in Advanced Economies

Standard international poverty lines are calibrated to low- and middle-income contexts; applied to wealthy economies, they register almost no one. Yet a family that earns above those thresholds and still cannot afford food, heating, or healthcare is deprived by any standard its own society would recognize. This section documents that deprivation using each country's own official measures, not to relabel these households under an international line they sit well above, but to show a form of insufficiency that those lines were never built to detect.

**Near-poverty in the United States.** The United States measures poverty through both an Official Poverty Measure (OPM) and the more comprehensive Supplemental Poverty Measure (SPM):

***Table 5. United States poverty by measure, 2024.***

| Measure | People (2024) | Rate |
|---|---|---|
| Official Poverty Measure | 35.9M | 10.6% |
| Supplemental Poverty Measure | 43.7M | 12.9% |
| Near-poverty (below 125% OPM) | ~50M | ~15% |
| Below 200% of poverty threshold | ~93.6M | ~30% |

Some 93.6 million Americans — close to 30% of the nation — live in households below twice the poverty threshold (U.S. Census Bureau, 2025; Center for American Progress, 2025). A family of four at this level earns less than about $62,400 (200% of the 2024 federal poverty guideline of $31,200). In high-cost metro areas, this places a family in chronic precarity: unable to save, unable to absorb a medical emergency. They are invisible in global statistics because they earn well above any international line.

**OECD relative poverty.** Across the 38 OECD member countries — roughly 1.4 billion people — the average relative poverty rate (below 50% of national median disposable income) was 11.4% in 2021 (OECD, 2024):

***Table 6. Relative poverty in selected OECD countries.***

| Country | Relative poverty rate |
|---|---|
| Costa Rica | 21.0% |
| United States | 18.1% |
| Israel | 16.9% |
| Japan | 15.4% |
| South Korea | 14.9% |
| Canada | 12.2% |
| Germany | 11.6% |
| France | 8.7% |
| Netherlands | 7.0% |
| Czech Republic | 5.5% |

Approximately 160 million people live in relative poverty across OECD nations. This figure is descriptive of the OECD and is not the same as the disjoint high-income population used in Section 13's composite estimate: the OECD includes middle-income members (Mexico, Colombia, Costa Rica, Chile, Türkiye) whose relatively poor are largely already captured within the $8.30/day count, so the two are not simply additive. Chronic relative deprivation nonetheless correlates with worse health, shorter life expectancy, higher rates of mental illness, and reduced social mobility in wealthy countries as much as in poor ones.

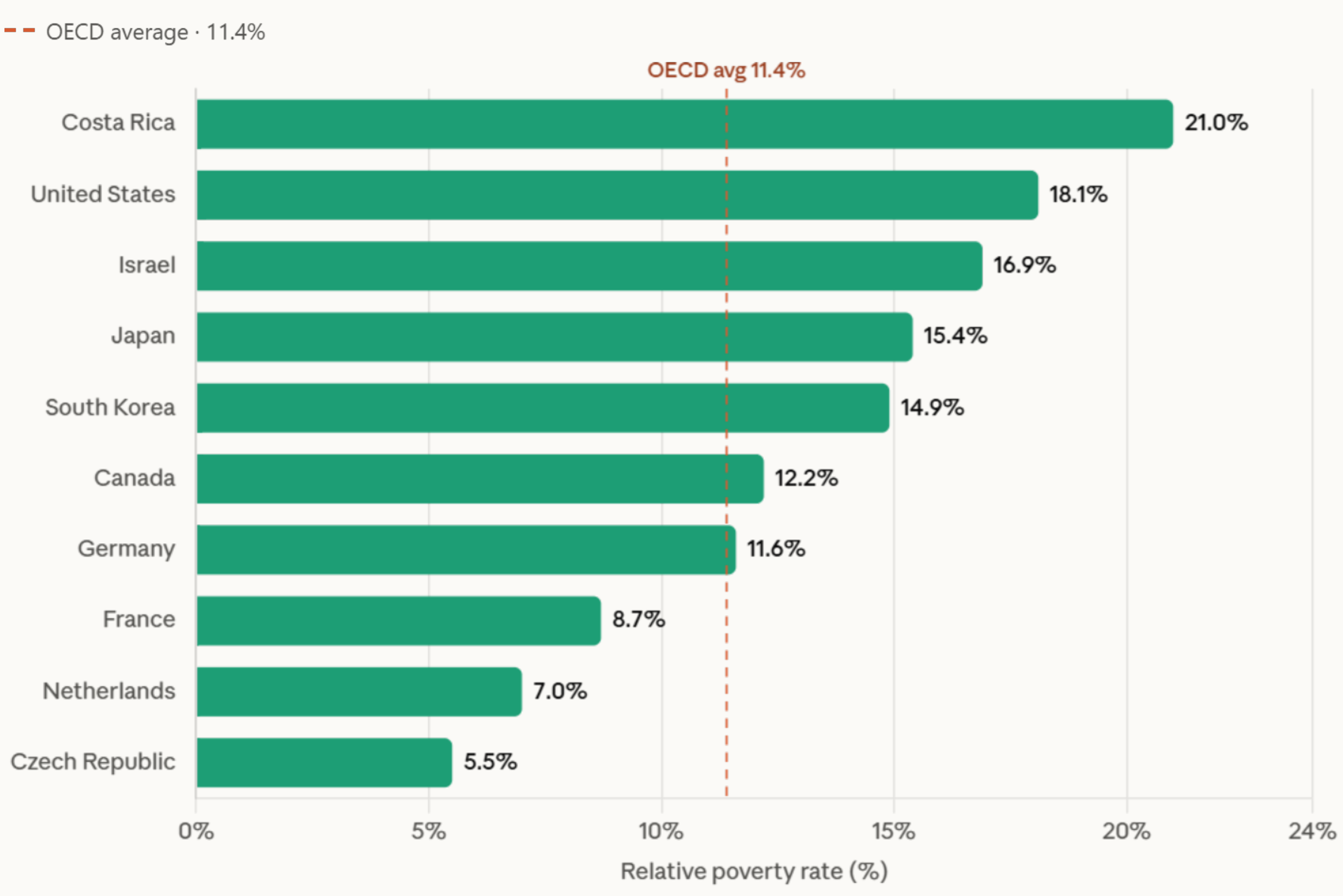

***Figure 4. Relative poverty across the OECD, 2021.***
*Source: OECD, Society at a Glance (2024).*

**The global "vulnerable" band.** The World Bank's Latin America and Caribbean analysis offers the clearest documented example of near-poverty (World Bank LAC, 2024): 31.5% of the region — about 211 million people — lived on $6.85–$14 per day in 2023, a population the Bank classifies as vulnerable and flags as "at major risk of falling into poverty during downturns or shocks." In the Caribbean, the share is higher, at 36% (World Bank Caribbean, January 2026). No comprehensive global figure for this band has been published, but combining the LAC data with known patterns in South Asia, Sub-Saharan Africa, and MENA suggests it very likely exceeds 800 million to 1 billion people worldwide — all counted as non-poor in official statistics.

**Child poverty and food deprivation.** In Italy, 13.8% of minors — 1.283 million children — lived in absolute poverty in 2024, the highest value recorded since 2014; among families with three or more minor children, the rate was 19.4%, and 2.7 million Italians (4.6%) experienced severe material or social deprivation (ISTAT, 2024). Across the European Union, 13.6% of children under 16 were materially deprived in 2024, with the burden concentrated in Greece (33.6%), Romania (31.8%) and Bulgaria (30.4%) (Eurostat, June 2025); a Save the Children analysis placed 19.5 million EU children — one in four — at risk of poverty or social exclusion, 446,000 more than in 2019 (ReliefWeb, October 2025). In the United States, 14.1 million children lived in food-insecure households in 2024 (10.1% of children in households with children), with 18.4% of households with children affected; the racial gradient is stark, at 31% for Black non-Hispanic households and 11.6% for Hispanic households (USDA, 2024).

Japan is the most paradoxical case. With GDP per capita around $33,000 and the highest single-parent employment rate in the OECD (86%), it still records child poverty of 11.5% — about two million children and a single-parent household poverty rate of 44.5%, among the highest in the OECD, reflecting structural wage floors rather than employment behavior (Japan MHLW 2022 Comprehensive Survey, via ISVD). Grassroots children's cafeterias grew from 319 in 2016 to more than 10,800 in 2024, surpassing the number of public junior high schools; Save the Children Japan reports that 90% of low-income households face rice shortages. Japan allocates roughly 15% of GDP to elderly social protection but only 1–2% to child support. In the United Kingdom, 4.5 million children — about 30% — were in poverty, rising to 36% where the youngest child is under five (Food Foundation, May 2025). Globally, a separate Save the Children estimate finds that 1.12 billion children — 48% of all children — cannot afford a healthy diet (March 2025), spanning both developing and advanced economies; it is a child-specific affordability measure distinct from the population-wide affordability figure discussed in Section 8.

**Energy poverty.** In 2024, 9.2% of the EU population reported being unable to keep their homes adequately warm, ranging from 19.0% in Bulgaria and Greece and 18.0% in Lithuania, down to 2.7% in Finland (Eurostat, February 2026). In Southern and Eastern Europe, energy poverty serves as a proxy for poverty that income statistics miss.

**Housing cost burden.** In eight OECD countries, more than 40% of low-income renters spend over 40% of their disposable income on housing; in the United States, Chile, Greece, and New Zealand, median housing costs exceed 40% of bottom-quintile disposable income; and in 14 countries, more than half of low-income tenants are overburdened once total housing costs are counted (OECD Affordable Housing Database, 2024). Greece is the extreme case, where 94% of low-income outright homeowners still face housing-cost overburden. A household spending more than 40% of its income on housing has, by definition, less than 60% for everything else.

**Medical debt (United States).** In 2024, 31 million Americans borrowed $74 billion for medical expenses (West Health/Gallup), and between 17.8% and 35% of adults carry medical debt at any given time (CFPB, 2022). Medical debt — the leading cause of personal bankruptcy in the United States — is a form of poverty that no income measure captures: a household may sit above the poverty line in the year it incurs catastrophic costs, while those costs shape its trajectory for years.

**The analytical point.** GDP per capita describes the average quantity of resources in an economy; it says nothing about their distribution or whether they translate into basic living standards at the lower end. Standard international lines, designed for settings where modest income growth is transformative, produce false negatives at scale when applied to advanced economies. The deprivation documented here — measured by each country's own definitions — is real, large, and systematically invisible in global poverty discourse.

## 8. Food and Nutrition

In 2024, 673 million people — 8.2% of the global population — faced chronic hunger (FAO SOFI, 2025; WHO, 2025). The geography maps onto poverty: Africa: 307 million (20.2%, rising); Western Asia: 39 million (12.7%, rising); Asia overall: 6.7%; Latin America and the Caribbean: 5.1%.

Hunger captures only the most severe end of deprivation. Food insecurity — the lack of reliable access to sufficient, nutritious food — affects 2.3 billion people (28%), 335 million more than in 2019 and 683 million more than in 2015 (FAO SOFI, 2025). By a related affordability measure, 2.6 billion people cannot afford a healthy diet (WHO, 2025) — a population that overlaps substantially with the food-insecure count above rather than adding to it. Both capture the same underlying reality from different angles: people whose income may exceed an official line but who cannot purchase the nutrients a human body requires.

## 9. Working Poverty and Social Protection

In 2024, more than 240 million workers — 6.9% of the global employed population — lived on less than $2.15/day (2017 PPP; UN SDG Report, 2025), and the ILO's 2026 employment report estimated that nearly 300 million workers were in some form of extreme working poverty. Working poverty is concentrated in the least developed countries (nearly 3 in 10 workers), landlocked developing countries (2 in 10), and small island developing states (9.5%, rising). It reflects structural labor-market conditions — wage floors, informality, bargaining coverage — rather than individual behavior, evidenced by the inverse correlation between social-protection spending and working-poverty rates across comparable economies. Beyond the working poor, a

further 402 million people face a global employment gap — a distinct population comprising 186 million unemployed, 137 million discouraged workers, and 79 million unable to work (ILO, 2025).

On social protection, 2023 marked the first year in which more than half the world's population — 52.4% — had access to at least one benefit, up from 42.8% in 2015 (UN SDG Report, 2025). The corollary is that 3.8 billion people remain entirely unprotected. Not all are poor, but the lack of protection sharply deepens the risk of poverty when illness, job loss, or climate shocks strike.

***Table 7. Social protection coverage by country income group, 2023.***

| Country income group | Social protection coverage |
|---|---|
| High-income | 85.9% |
| Upper-middle-income | 71.2% |
| Lower-middle-income | ~47% (estimated) |
| Low-income | 9.7% |

High-income countries spend an average of 24.9% of GDP on social protection; low-income countries spend 2.0%. Closing this gap would require an estimated additional $1.4 trillion per year.

## 10. Structural Amplifiers: Inequality and Inflation

**Inequality.** Higher inequality compounds poverty at any given level of national income — one of the most robust relationships in development economics. Two countries with the same average income but different Gini coefficients will have systematically different poverty rates (Gini reported here on the 0–1 scale):

- The United States (Gini ~0.49 for households, 2024) has a higher relative poverty rate (18.1%) than France (Gini ~0.33; 8.7%) or Germany (Gini ~0.32; 11.6%) despite higher GDP per capita. The difference is distributional, not aggregate.
- Colombia (Gini ~0.54) and Honduras (Gini ~0.47) show poverty rates of 31.8% and 62.9%, respectively — rates that reflect inequality compounding poverty beyond what income levels alone would predict.
- Denmark (Gini ~0.29) has a relative poverty rate of 6.3%. The United States, with a GDP per capita roughly a quarter higher than Denmark's, has a relative poverty rate nearly three times higher. The gap is explained by distribution, not wealth.

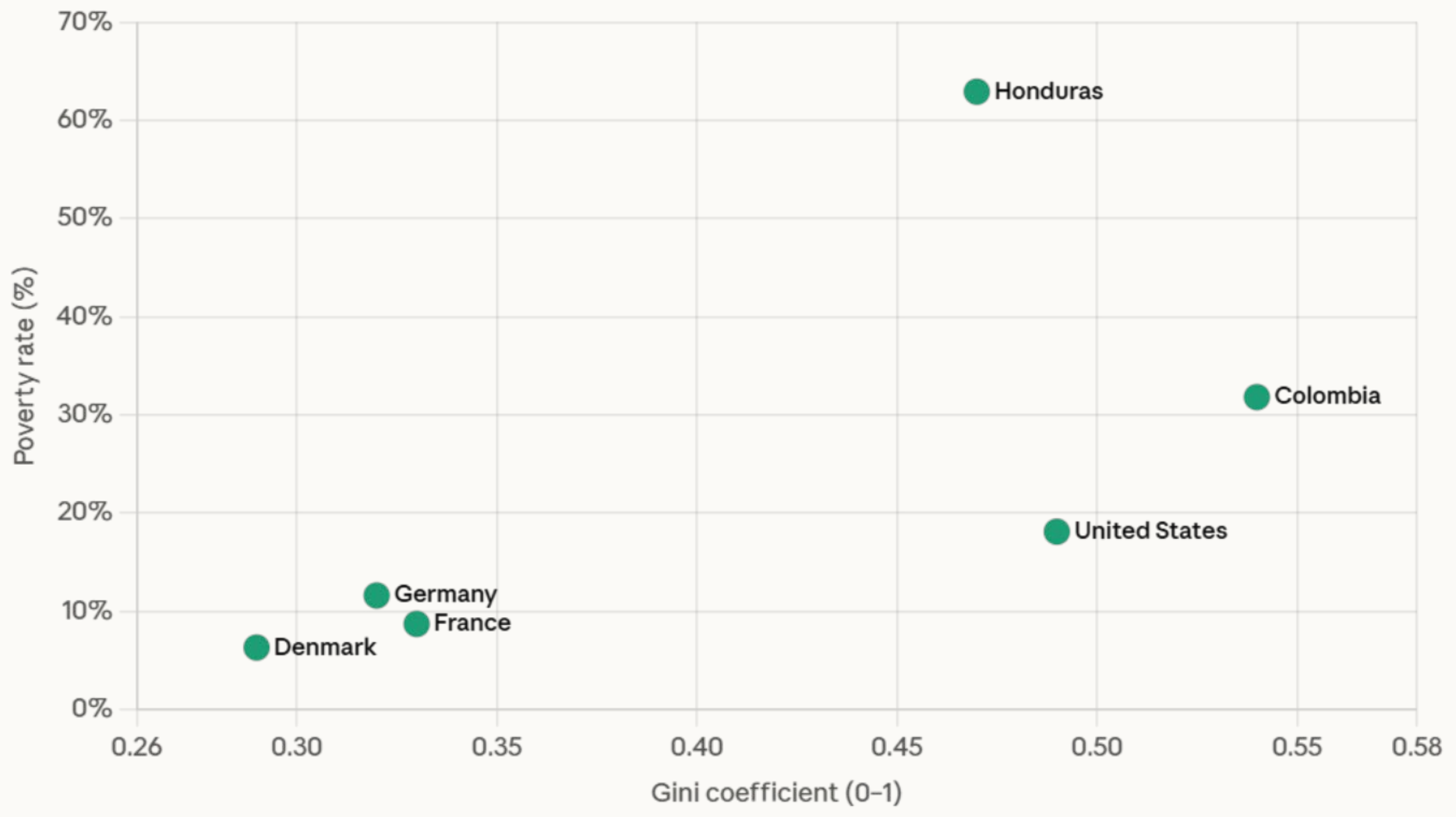

***Figure 5. Inequality compounds poverty.***
*Sources: OECD, Society at a Glance (2024); U.S. Census Bureau and FRED; World Bank World Development Indicators.*

Research finds that a 1% increase in social transfers as a share of household income reduces income inequality by about 0.13%, while increases in poverty itself raise inequality by about 0.34%, a reinforcing feedback loop (Kerimov, 2024). Poverty and inequality amplify one another.

**Inflation.** Poor households spend a different basket of goods than wealthy ones: across most economies, lower-income families allocate 50–80% of expenditure to food, fuel, and housing — precisely the categories that inflated most sharply in 2021–2024. Brookings (2024) finds that standard CPI overstates the purchasing-power gains of poor families by 18–25% relative to a necessities-based index over the past fifty years. When US inflation was recorded at 8–9% in 2022, the effective rate for bottom-quintile households was materially higher. The global dimension is larger still: Pakistan's full-year headline CPI averaged 29.2% in FY23 and 23.4% in FY24, peaking at 37.97% in May 2023 (IMF, September 2024); Brazil reached 13.43% food inflation in 2022; and several Sudanese, Yemeni, and Sub-Saharan economies saw food prices rise 30–50%. During inflationary periods, poverty measured against nominal income thresholds understates real deprivation.

## 11. The Gender Dimension

Gendered poverty is structural, not a simple percentage-point gap. Multiple reinforcing mechanisms are evident. Women in Latin America and the Caribbean face a 42% higher risk of poverty than men among young adults, driven by lower labor-force participation, worse job quality, and unpaid care responsibilities (World Bank, Fall 2025). Single-mother households constitute roughly 25% of poor households and contain 31 million people in extreme poverty — a higher poverty risk than any other household type. Intra-household allocation systematically disadvantages women and girls in food, education spending, and healthcare, a dimension absent from income-based measures. Regarding property rights, only 24% of adult women hold legal land documents, compared with 43% overall. In Sub-Saharan Africa, the gender poverty gap accounts for three-quarters of the global total of women in extreme poverty, compounded by substantially lower social-protection access. A count that does not disaggregate by gender will underrepresent the severity of deprivation among women.

## 12. Trajectory: Progress, Stagnation, and Reversal

Between 1990 and 2019, the global extreme-poverty rate fell from roughly 36% to under 9% — one of the most significant reductions in human deprivation on record, driven primarily by growth in East Asia, especially China. This progress was real and should not be minimized.

The post-2019 record is less favorable. COVID-19 (2020–2021) reversed an estimated five years of progress, pushing 70–90 million additional people into extreme poverty. Conflict has intensified, with 451 million people in fragile and conflict-affected states now in extreme poverty, and the population of such states has more than doubled since 2005. Climate shocks increasingly displace the rural poor in Sub-Saharan Africa and South Asia; debt crises constrain fiscal space for social protection; and humanitarian assistance fell 7.1% in 2024, with a further 9–17% decline projected for 2025 (Global Hunger Index, 2025).

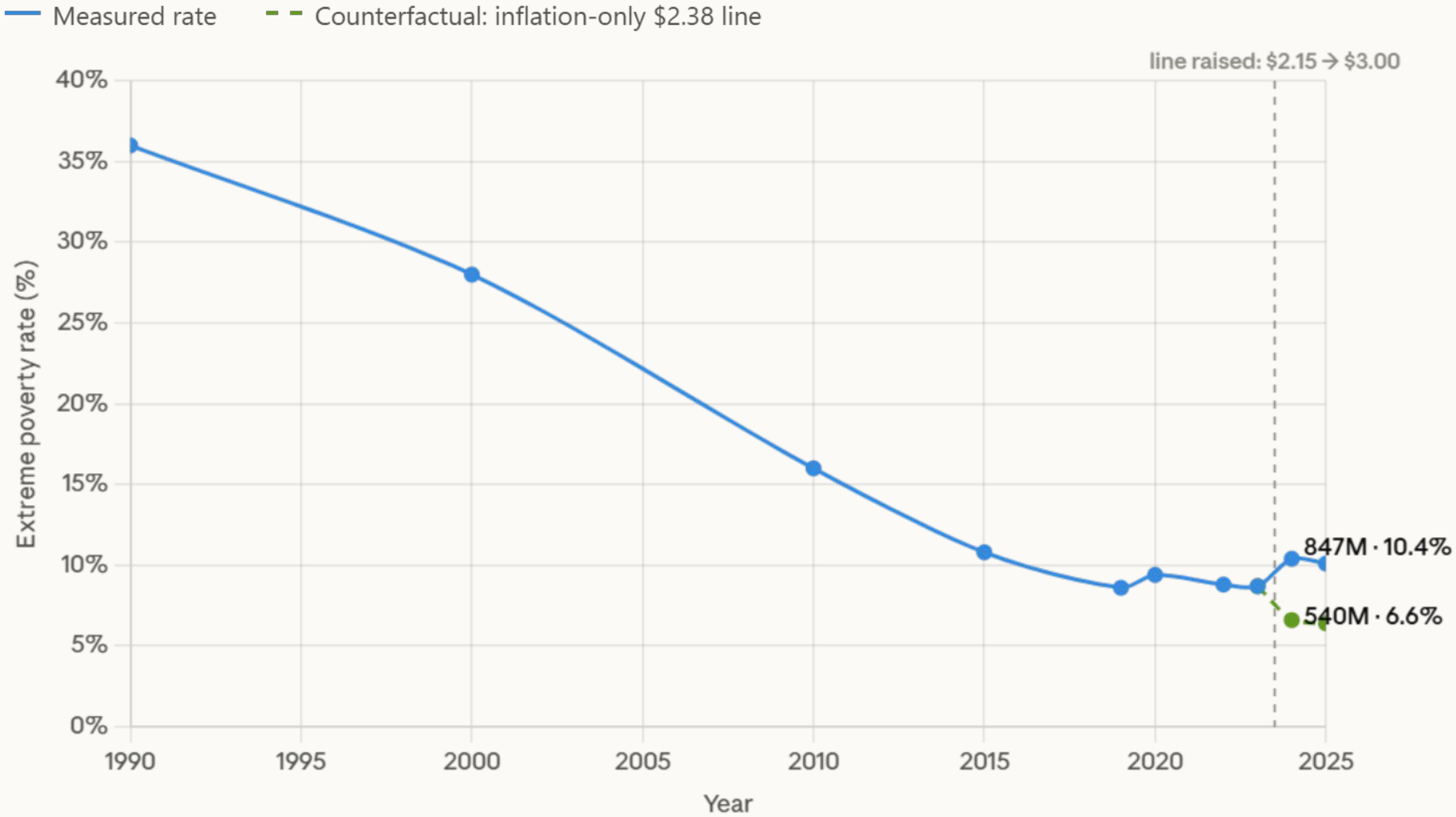

***Figure 6. Global extreme poverty rate, 1990–2025, with a post-2024 threshold break.***
*Source: World Bank PIP (March 2026); counterfactual per Our World in Data (August 2025)*

The measured rate fell from roughly 36% to under 9% before rising in 2024, when the International Poverty Line was raised from $2.15 (2017 PPP) to $3.00 (2021 PPP). The dashed counterfactual shows that on an inflation-only line (~$2.38), 2024 would read about 540 million rather than the headline 847 million — most of the apparent rise reflects the higher line, not worsening conditions

## 13. Constructing a Bounded Estimate

The count depends entirely on the standard applied. This report presents three estimates, each grounded in the data but reflecting different — and clearly distinguished — analytical choices. They are not competing measures of a single true number; they answer different questions.

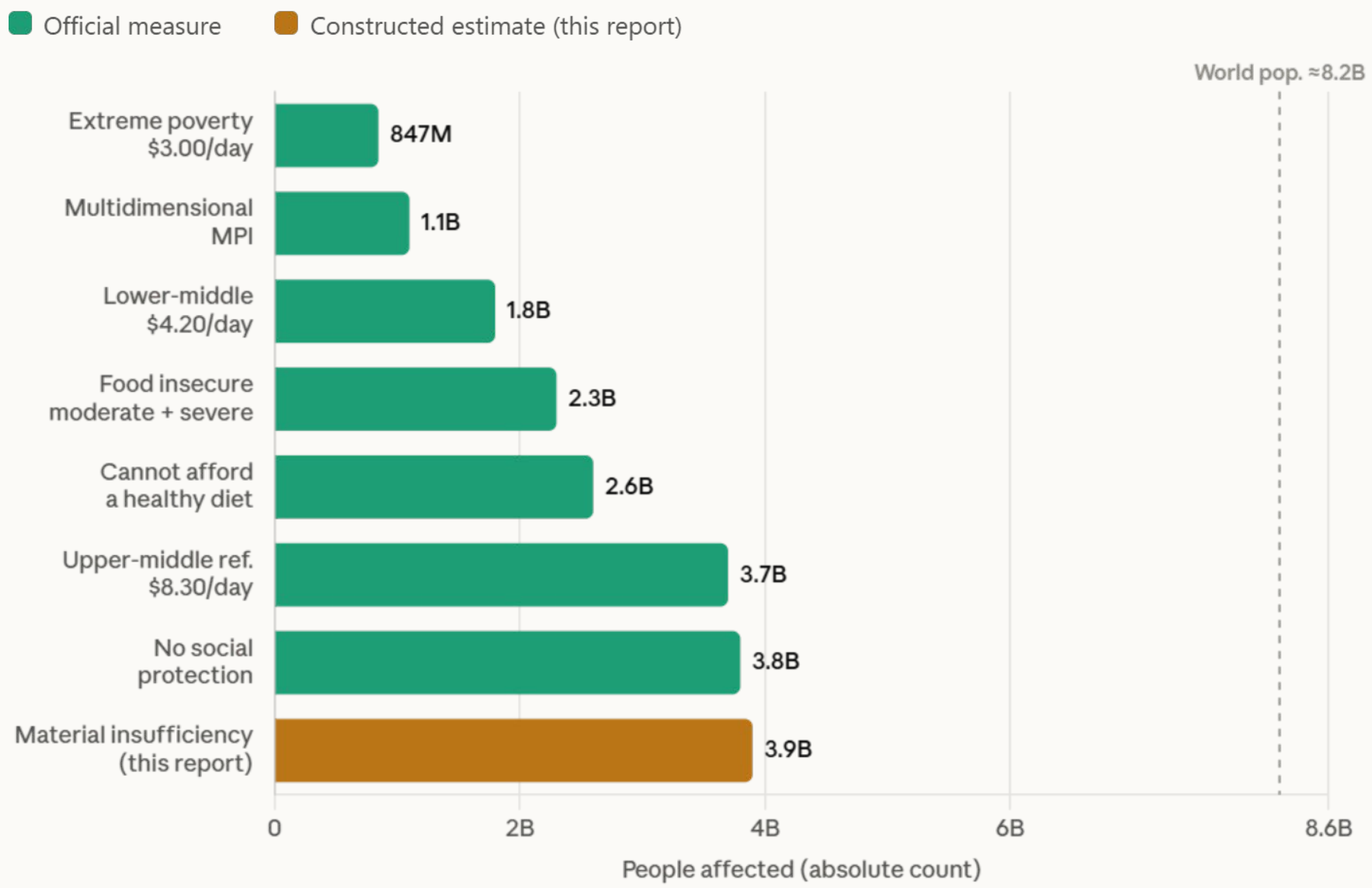

***Figure 7. The global poverty count is standard-dependent.***
*Sources: World Bank PIP (March 2026); UNDP/OPHI (2024); FAO SOFI (2025); ILO/UN SDG (2023); author's estimate*

**Full metric landscape.** The categories below overlap: a person may be simultaneously food insecure, multidimensionally poor, a working-poor adult, and without social protection. The table's purpose is to convey scope across dimensions, not to be summed.

*Table 8. The full metric landscape: overlapping measures of poverty and deprivation.*

| Metric | People | % of world | Certainty | Source |
|---|---|---|---|---|
| Extreme poverty ($3/day) | ~847M | 10.4% | High | World Bank PIP, Mar 2026 |
| Multidimensional poverty | 1.1B | 13.5% | High | UNDP/OPHI, 2024 |
| Lower-middle poverty ($4.20/day) | ~1.8B | ~22% | Medium | World Bank PIP, 2025 (approx.) |
| Food insecure (moderate + severe) | 2.3B | 28.0% | High | FAO SOFI, 2025 |
| Cannot afford a healthy diet | 2.6B | 31.9% | High | FAO SOFI / WHO, 2025 |
| Upper-middle poverty ($8.30/day) | 3.7B | 45.5% | Medium-High | World Bank, Fall 2025 |
| Without social protection | 3.8B | 47.2% | High | ILO / UN SDG, 2023 |
| Working poor (<$2.15/day, 2017 PPP) | 240M+ | 6.9% of employed | High | UN SDG Report, 2025 |
| Relative poor in OECD | ~160M | 11.4% of OECD | Medium | OECD, 2024 |
| Vulnerable / near-poor (global est.) | ≥800M–1B | ~10–12% | Low-Medium | LAC documented; global extrapolated |

**Three bounded estimates.** The high-income relative-poor component used below (~150–200M) is the author's estimate, derived from national relative-poverty counts in high-income countries net of the population already below $8.30/day in PPP terms. It is distinct from the ~160M OECD figure in the table above, which includes middle-income OECD members already captured by the $8.30 line.

*Conservative floor — approximately 1.0–1.2 billion.* Extreme monetary poverty (~847M) plus the disjoint set of high-income-country relative poor above $8.30/day PPP (~150–200M). This uses only the most conservative monetary thresholds and excludes near-poverty and non-monetary deprivation. *(Certainty: high.)*

*Central estimate — approximately 3.85–3.90 billion.* People below the $8.30/day upper-middle-income reference line (~3.7B) plus the disjoint high-income relative-poor set (~150–200M). This is an analytical construct, not a World Bank poverty count. It operationalizes a specific normative standard — insufficiency relative to the norms of one's own society — using the upper-middle-income line as a global anchor and supplementing it only with the high-income population that the line does not reach. The two populations are essentially non-overlapping. Readers who reject the anchor should treat the conservative floor as the headline. *(Certainty: medium; sensitive to the choice of anchor.)*

*Upper bound — approximately 4.5–5.0 billion.* The central estimate plus a globally extrapolated vulnerable band (~800M–1B), documented directly only for Latin America and the Caribbean. *(Certainty: low; extrapolated.)*

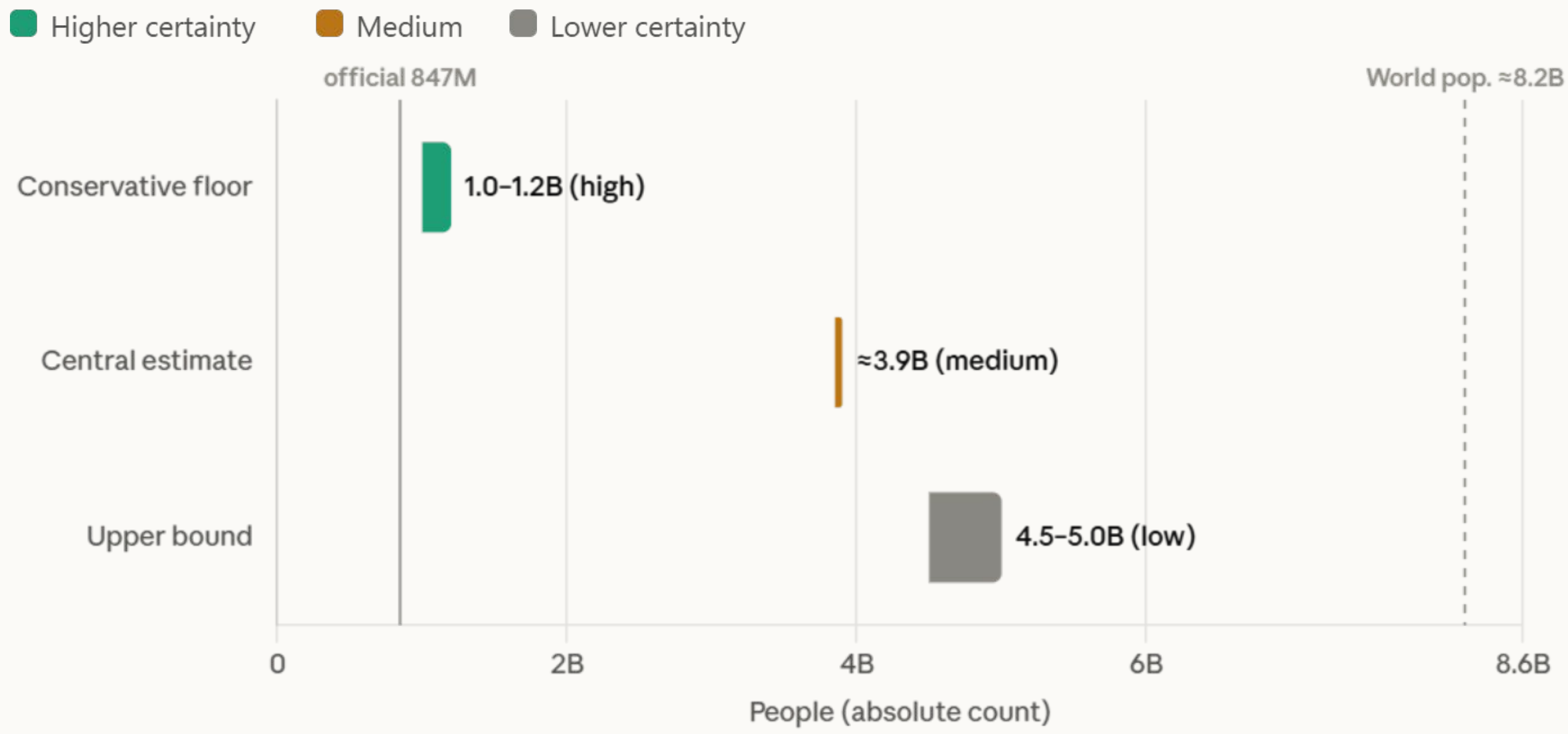

***Figure 8. Three bounded estimates.***
*Source: author's estimates, constructed from World Bank, UNDP/OPHI, FAO, OECD, and ILO data.*

The width of this range is the finding. The honest answer to "how many people live in poverty?" is not a single number but a function of the standard chosen — and that function runs from roughly one billion to roughly five.

## 14. Recommendations

The evidence in this report supports methodological recommendations before they are political, because measurement failure is the most consistent thread running through it.

First, close the survey gap at its widest point. Fewer than half of countries have post-2020 data, and coverage is thinnest in the highest-poverty regions. Concessional financing for national statistical systems in fragile and conflict-affected states, and specifically in Sub-Saharan Africa, should be treated as poverty-reduction infrastructure rather than administrative overhead, and ring-fenced against the humanitarian funding contractions now underway.

Second, standardize the reporting of thresholds and price bases. Every published poverty figure should state its line, and its PPP vintage, and cross-vintage comparisons should be flagged rather than presented as continuous series. The 2023/24 break in the extreme-poverty trend is a case study in how an unlabeled methodological change can be misread as a change in the world.

Third, report deprivation as a range with graded certainty rather than a single headline. Institutions already hold the components; presenting the $3.00, $8.30, and multidimensional figures together, each with its certainty and coverage, would directly communicate the standard-dependence of the count and reduce the risk that the narrowest measure stands in for the whole.

Fourth, mandate disaggregation by age, gender, and, where feasible, intra-household allocation. Children and women are the populations most systematically undercounted by household-level monetary measures, and the tools to disaggregate exist.

Fifth, publish a defined "vulnerable" or near-poverty band beyond the poverty line itself. The Latin America and Caribbean work demonstrates the value of quantifying the population within one shock of poverty; extending it globally would make visible a population that all headline statistics omit.

## Conclusion

The headline figure — approximately 847 million people, 10.4% of the world — is accurate, and this report treats it as such. The argument is not that the number is wrong but that it is narrow: it marks the edge of survival in the poorest economies and has come to stand in for a question it was never built to answer. Asked the fuller question — how many people cannot reliably meet basic needs by the standards of their own society — the data support no single figure, but they do support a range whose floor already exceeds a billion and whose central case approaches half of humanity.

Three points survive the measurement uncertainty. The count is standard-dependent and spans an order of magnitude, so any figure presented without its threshold is misleading. The highest-poverty regions have the weakest measurement infrastructure, so official figures are likely to undercount rather than overcount. And deprivation in advanced economies — food-insecure children, working single parents below national poverty lines, households in energy or medical-debt distress — is real and measurable by those countries' own standards, yet invisible to every international line in use.

That is not a claim engineered for effect; it is what follows from taking every available measure seriously and stating the assumptions behind each. Progress from 1990 to 2019 was genuine and substantial and should not be minimized. But it has stalled, and in the poorest regions, it has been partly reversed, and the pace is no longer adequate to meet the target the world set for 2030. The appropriate response is methodological before it is rhetorical: better surveys, more frequent collection, consistent treatment of thresholds and price bases, and honest reporting of uncertainty.

The scale of deprivation has been underrepresented in public discourse less by dishonesty than by the quiet dominance of the most conservative available measure. Naming that clearly — and holding the fuller range, uncertainty, and all — is the first step toward measuring, and then addressing, the problem as it actually is.

## References

World Bank, *Poverty and Inequality Platform (PIP), March 2026 Global Poverty Update.* https://blogs.worldbank.org/en/opendata/march-2026-global-poverty-update-from-the-world-bank--new-data-a

World Bank, *PIP September 2025 Global Poverty Update.* https://blogs.worldbank.org/en/opendata/september-2025-global-poverty-update-from-the-world-bank--new-da

World Bank, *Poverty and Inequality Update, Fall 2025.* https://thedocs.worldbank.org/en/doc/229ff18129687a785f08af7cfb28e5e1-0350012025/original/WBG-Poverty-and-Inequality-Update-Fall-2025.pdf

World Bank, *International Poverty Line Revision, June 2025.* https://documents.worldbank.org/en/publication/documents-reports/documentdetail/099510306052516849

World Bank, *Poverty and Inequality Platform (PIP).* https://pip.worldbank.org

World Bank, *Poverty Projections for Pakistan.* https://documents1.worldbank.org/curated/en/099459512302438604/pdf/IDU17bf933d2197611440918caf1f63cb8d5bfdc.pdf

World Bank, *South Asia Blog: Pakistan's Poverty Trajectory, September 2025.* https://blogs.worldbank.org/en/endpovertyinsouthasia/pakistan-s-poverty-trajectory--progress--peril--and-the-path-for

World Bank, *Pakistan Development Update, April 2024.* https://thedocs.worldbank.org/en/doc/140b30353b40dbb294cca42bcb86529a-0310062024/original/Pakistan-Development-Update-April-2024.pdf

World Bank, *Filling the Gaps in Survey Data (PIP Coverage Blog), November 2024.* https://blogs.worldbank.org/en/opendata/filling-the-gaps-in-survey-data-for-a-world-free-of-poverty-on-a

World Bank / Universidad de Los Andes, *Survey Gaps Leave Poverty Predictions on Shaky Ground, August 2025* (reported by Devdiscourse). https://www.devdiscourse.com/article/other/3570387-survey-gaps-leave-poverty-predictions-on-shaky-ground-world-bank-researchers-say

World Bank, *Poverty in Latin America: 10 Facts for 2024.* https://blogs.worldbank.org/en/opendata/poverty-in-latin-america--10-facts-you-need-to-know-for-2024

World Bank, *Solving Caribbean Poverty Starts With Seeing It, January 2026.* https://www.worldbank.org/en/country/caribbean/brief/solving-caribbean-poverty-starts-with-seeing-it-new-data-reveal-hidden-vulnerabilities

World Bank (De Vreyer and Lambert), *Data Gaps: The Poor Typical Household Surveys Miss.* https://blogs.worldbank.org/en/africacan/data-gaps-the-poor-typical-household-surveys-miss

IMF, *Article IV Consultation: Pakistan, September 2024.* https://www.imf.org/en/news/articles/2024/09/27/pr-24343-pakistan-imf-concludes-2024-aiv-consultation-pakistan-approves-37-mo-extended-arr

UNDP/OPHI, *Global Multidimensional Poverty Index, 2024.* https://hdr.undp.org/content/2024-global-multidimensional-poverty-index-mpi

OPHI. https://ophi.org.uk/

FAO, *State of Food Security and Nutrition in the World, 2025.* https://openknowledge.fao.org/handle/20.500.14283/cd9562en

WHO, *Global Hunger Declines but Rises in Africa and Western Asia, July 2025.* https://www.who.int/news/item/28-07-2025-global-hunger-declines-but-rises-in-africa-and-western-asia-un-report

ILO, *World Employment and Social Outlook, 2025.* https://dsv-europa.de/en/news/2025/02/ilo-report-2025.html

United Nations, *SDG Report 2025, Goal 1.* https://unstats.un.org/sdgs/report/2025/goal-01/

OECD, *Society at a Glance, 2024 — Income Poverty.* https://www.oecd.org/en/publications/society-at-a-glance-2024_918d8db3-en/full-report/income-poverty_53d4eac1.html

OECD, *Affordable Housing Database HC1.2, 2024.* https://webfs.oecd.org/els-com/Affordable_Housing_Database/HC1-2-Housing-costs-over-income.pdf

Our World in Data, *The New International Poverty Line of $3 a Day, August 2025.* https://ourworldindata.org/new-international-poverty-line-3-dollars-per-day

U.S. Census Bureau, *Income in the United States: 2024.* https://www.census.gov/library/publications/2025/demo/p60-286.html

U.S. Census Bureau, *Poverty in the United States: 2024.* https://www.census.gov/library/publications/2025/demo/p60-287.html

Center for American Progress, *Poverty in America Remained Flat in 2024, 2025.* https://www.americanprogress.org/article/poverty-in-america-remained-flat-in-2024-but-will-likely-rise-as-the-one-big-beautiful-bill-act-goes-into-effect/

Federal Reserve Bank of St. Louis (FRED), *Gini Index (United States).* https://fred.stlouisfed.org/series/GINIALLRF

Federal Reserve Bank of St. Louis (ALFRED), *Gini Index (Germany).* https://alfred.stlouisfed.org/series?seid=SIPOVGINIDEU

ISTAT, *Poverty Statistics, Year 2024.* https://www.istat.it/en/press-release/istat-poverty-statistics-year-2024/

Eurostat, *Child Material Deprivation, June 2025.* https://ec.europa.eu/eurostat/web/products-eurostat-news/w/ddn-20250613-2

Eurostat, *Energy Poverty, February 2026.* https://ec.europa.eu/eurostat/web/products-eurostat-news/w/ddn-20260202-2

Save the Children / ReliefWeb, *Child Poverty: A Cost Europe Cannot Afford, October 2025.* https://reliefweb.int/report/world/child-poverty-cost-europe-cannot-afford

Save the Children, *Healthy Diet Now Unaffordable for Nearly Half the World's Children, March 2025.* https://www.savethechildren.net/news/healthy-diet-now-unaffordable-nearly-half-worlds-children-save-children-analysis

USDA, *Household Food Security in the United States in 2024.* https://ers.usda.gov/sites/default/files/_laserfiche/publications/113623/ERR-358.pdf

Japan Ministry of Health, Labour and Welfare, *2022 Comprehensive Survey* (via ISVD). https://isvd.or.jp/en/columns/2026-03-09-child-poverty-depth-japan

Food Foundation (UK), *Price a Barrier to Feeding Children, May 2025.* https://foodfoundation.org.uk/press-release/more-third-parents-say-price-barrier-being-able-feed-their-child-what-theyd

Consumer Financial Protection Bureau, *Medical Debt Burden in the United States, 2022.* https://files.consumerfinance.gov/f/documents/cfpb_medical-debt-burden-in-the-united-states_report_2022-03.pdf

Global Hunger Index, *2025.* https://www.globalhungerindex.org/pdf/en/2025.pdf

Brookings Institution, *The Cost of Being Poor Is Rising, 2024.* https://www.brookings.edu/articles/the-cost-of-being-poor-is-rising-and-its-worse-for-poor-families-of-color/

Kerimov, A., *Business Perspectives, 2024.*
https://www.businessperspectives.org/images/pdf/applications/publishing/templates/article/assets/21172/PMF_2024_02_Kerimov.pdf

## About This Report

This report is an independent evidence synthesis on the measurement of global poverty, prepared by the Institute for Machine Learning in collaboration with the AI Institute at St. John's University and published by the Institute for Machine Learning. It draws entirely on publicly available data from international institutions and national statistical agencies, cited in full in the References.

It is offered as a free public download. Readers are encouraged to consult the primary sources directly and to reach the corresponding author with questions or corrections.

*Institute for Machine Learning · Danbury, Connecticut*
*AI Institute, St. John's University · Queens, New York*
*Correspondence: crocettg@stjohns.edu*

# About the authors

## Giancarlo Crocetti

Assistant Professor, St. John's University · corresponding author

As the director of the AI Institute @ SJU, he focuses on the impact of AI on society and pedagogy. Before joining St. John's, he was a Senior Associate Director at Boehringer Ingelheim Pharmaceuticals for more than 18 years, leading the Data Science & Advanced AI capability for the medical organization. He is now focusing on applying Generative AI to prepare students and professionals for success in the future workforce. He also collaborates with the Institute for Machine Learning to find novel ways to use Generative AI in line with the University's Vincentian mission, for the benefit of all.

crocettg@stjohns.edu

## Editorial Team

Institute for Machine Learning · contributing authors and publisher

A multidisciplinary, international group of graduate students at the Institute for Machine Learning. Working across disciplines and countries, they carried out the research, analysis, and review behind this report.

## About the institutions

AI Institute, St. John's University · Queens, NY

The AI Institute supports research and education in artificial intelligence and machine learning at St. John's University.

Institute for Machine Learning · Danbury, CT

An independent research organization advancing machine learning in the service of the public good.

### Contact and full report

Corresponding author: crocettg@stjohns.edu

Read the full report, free at www.i4ml.com